\documentstyle[epsfig]{aa}
\def\beq{\begin{equation}}
\def\enq{\end{equation}}
\def\bea{\begin{eqnarray}}
\def\ena{\end{eqnarray}}
\def\bec{\begin{center}}
\def\enc{\end{center}}

\def\ergcm2si{\hbox{ergs~cm$^{-2}$s$^{-1}$}}

\begin{document}

\title{Deprojected analysis of Abell 1835 observed with {\it\bfseries Chandra} and compared with {\it\bfseries XMM-Newton}}

\author{ C.K. Li\inst{1}, S.M. Jia\inst{1}, Y. Chen\inst{1}, F. Xiang\inst{2}, Y.S. Wang\inst{1}, \and H.H. Zhao\inst{1}}

\offprints{ Chengkui~Li (E-mail: lick@ihep.ac.cn)}

\institute{Key Laboratory for Particle Astrophysics, Institute of High Energy Physics,
Chinese Academy of Sciences, 19B Yuquan Road, Beijing 100049, China
\and
Department of Physics, Yunnan University, 650091, Kunming, China}

\date{2012.08 }

\abstract{Using a deprojection technique, we study the X-ray
properties of the galaxy cluster Abell 1835 observed with {\it Chandra}, including temperature, abundance, electron density, gas mass fraction, and total mass.
A comparison with the results without deprojection shows that the properties do not change much.
When we compare the results with those of {\it XMM-Newton}, the difference between the temperature profiles derived from {\it Chandra} and {\it XMM-Newton} data still exists, even if the point-spread function effect of {\it XMM-Newton} is corrected. To investigate the reasons for the difference, we used the spectra to cross-calibrate the temperatures. They show that the {\it Chandra} spectra can be fitted well with {\it XMM-Newton} temperatures.
Furthermore, we derive the electron density profile from {\it Chandra} data with {\it XMM-Newton} temperatures and calculate the projected mass,
which is consistent with the {\it XMM-Newton} mass and a little lower than the weak lensing mass at $r_{500}$. Thus, it seems that the temperature derived from {\it XMM-Newton} may be more reliable.
\\
}

\titlerunning{The deprojected analysis of Abell 1835 observed by {\it Chandra} and comparison with {\it XMM-Newton}}
\authorrunning{Li et al.}

\maketitle

%%%%%%%%%%%%%%%%%%%%%%%%%%%%%%%%%%%%%%%%%%%%%%%%%%%%%%%%%%%%%%%%%%%%%%%%%%%%%%%
\section{Introduction}
Galaxy clusters, the largest objects in the universe, originate in the primordial density perturbations from cosmic gravitational collapse.
 They are used in a variety of ways to perform both cosmological and astrophysical studies.
Modern astronomy satellites such as {\it XMM-Newton} and {\it Chandra} have high sensitivity and spatial resolution,
and unprecedented results have been gained through detailed analysis of their data. However, there are discrepancies between the properties of galaxy clusters
derived from {\it Chandra} and {\it XMM-Newton} data, such as gas temperature and total mass.
Nevalainen et al. (2010) examined the cross-calibration of the energy dependence and normalization
 of the effective area of {\it Chandra} and {\it XMM-Newton}, finding that the discrepancies of the $0.5-7.0$ keV band temperature measurements of
 galaxy clusters with {\it EPIC/XMM-Newton} and {\it ACIS/Chandra} could reach $\sim10-15\%$ on average.% It might be the dominated element for the discrepancy.

Abell 1835 is a classical bright cluster with a big cool core  (Allen et al. 1996). Its X-ray morphology (Schmidt et al. 2001; Smith et al. 2005)
shows that it is an undisturbed and relaxed cluster.
It is also an optimal candidate for a triaxial joint analysis via X-ray, SZ, and lensing techniques (Morandi et al. 2011).
The temperature difference between the {\it Chandra} and {\it XMM-Newton} analysis also exists in Abell 1835:
$\sim12$ keV from {\it Chandra} data (Schmidt et al. 2001)
and $\sim7.6$ keV from {\it XMM-Newton} data (Jia et al. 2004).
After the point-spread function (PSF) correction of {\it XMM-Newton} data, Wang et al. (2010) derived a temperature profile similar to that of Jia et al. (2004).
Therefore, the difference is not due to the PSF effect.

In addition, the masses of Abell 1835 derived from {\it XMM-Newton} and {\it Chandra} are different (Jia et al. 2004; Schmidt et al. 2002).
Fortunately, the masses have been measured with the strong lensing method (Richard et al. 2010) as well as with the weak lensing method (Zhang et al. 2008).
Since gravitational lensing directly probes the cluster total mass without any strong assumptions about the equilibrium state of the cluster,
the lensing mass is generally more reliable.

Recently, the comparison of X-ray and gravitational lensing masses has been studied in detail by both observational and simulated analyses.
Generally, the X-ray mass is consistent with or lower than the gravitational lensing mass.
Richard et al. (2010) showed that the ratio of strong lensing mass and X-ray projected mass
within $r<250$ kpc, $M_{SL}/M_X$, was 1.3.
It was also found that the mass derived from the X-ray measurement is about half of the strong lensing mass
in some clusters (e.g., A1689, Andersson \& Madejski 2004, Lemze et al. 2008; PKS 0745-191, Chen et al. 2003).
Moreover, Zhang et al. (2008) showed that the average ratio of the weak lensing mass to X-ray mass was $1.09\pm0.08$,
while Mahdavi et al. (2008) demonstrated that $M_X/M_{WL}$ is $1.03\pm0.07$ and $0.78\pm0.09$ at $r_{2500}$ and $r_{500}$, respectively.
N-body/hydrodynamical simulation work also estimated the ratio between X-ray and lensing masses,
 $M_X/M_{WL}$,  which was $0.88\pm0.02$ and $0.75\pm0.02$ at $r_{500}$ in Meneghetti et al. (2010)
and Rasia et al. (2012), respectively.

In this work, we reanalyse the {\it Chandra} data of Abell 1835 with the same deprojection technique as Jia et al. (2004)
to ascertain if the differences in temperatures derived from {\it Chandra} and {\it XMM-Newton} data can be corrected
by data analysis.
Furthermore, we investigate the reasons for the temperature differences and establish which result is more reasonable.

This paper is organized as follows:
The {\it Chandra} observations and data preparation are described in Section 2. Section 3 shows the basic spectra analysis, while Section 4 presents the deprojected electron density profile
and the mass profile. In Section 5, we discuss the reasons for the temperature differences derived from {\it Chandra} and {\it XMM-Newton} data. We present our conclusions in Section 6.

Throughout this paper, we assume $H_0$ = 70 km s$^{-1}${Mpc}$^{-1}$, $\Omega_\Lambda=0.7$, $\Omega_m=0.3$. One arcminute corresponds to 236.2 kpc at Abell 1835 redshift of 0.2523.
The selected energy band is 0.5 keV- 7.0 keV.

%%%%%%%%%%%%%%%%%%%%%%%%%%%%%%%%%%%%%%%%%%%%%%%%%%%%%%%%%%%%%%%%%%%%%%%%%%%%%%%
\section{Observation and spectra extraction}
Abell 1835 was observed with {\it Chandra} on 25 August 2006 for a total of 119.48 ksec (observation ID is 6880).
The observation instrument was ACIS-I, and the observation model was VFAINT.
We processed the {\it Chandra} data with CIAO 4.2 and CALDB 4.2.0.

%\subsection{Background correction}
Background images were extracted from the standard set of CTI-corrected ACIS blank sky images in the {\it Chandra} CALDB (Markevitch et al. 2003).
To remove the particle background, we estimated the count rates in the hard energy band  (10-12 keV) of Abell 1835 and the blank sky,  and  renormalized the blank sky.
We used the tool LC\_CLEAN in CIAO to scan the light curve of data for flares.
The Good Time Interval (GTI) was about 117 ksec.
%\subsection{Spectral deprojection}
Because Abell 1835 appears to be a relaxed cluster of galaxies, we assumed
that the temperature structure of this cluster is spherically symmetric and
applied a deprojection technique.
The deprojected spectrum of each shell is calculated by subtracting
the contributions from all the outer shells (e.g., Nulsen \& B\"{o}ringer 1995; Matsushita et al. 2002).
The detailed calculation procedures were described in Chen et al. (2003) and Jia et al. (2004, 2006).

We divided the image of the cluster into seven annular regions centered on the
emission peak for the extraction of spectra and used the outmost ring
($8.33'-10.42'$) to determine the local cosmic X-ray background (CXB).
 %The minimum width of the rings was set to $0.25'$.
For each annular region, ancillary response files (ARFs) and response matrix files (RMFs) are generated using CIAO.
The complete process is:
a) select GTI using the light curve of data; b) subtract the point sources; c) subtract the background using blank sky data;
d) extract the project spectra; e) extract the deprojected spectra.

For the purpose of comparing, we also reprocessed the {\it XMM-Newton} data of Abell 1835 (observation ID: 0551830201) in $\Lambda$CDM cosmology as done in Jia et al. (2004) using SAS 11.0.0.
%%%%%%%%%%%%%%%%%%%%%%%%%%%%%%%%%%%%%%%%%%%%%%%%%%%%%%%%%%%%%%%%%%%%%%%%%%%%%%%
\section{Spectral analysis}

For the spectral analysis, we used the plasma emission model MEKAL (Mewe et al. 1985, 1986; Kaastra 1992; Liedahl et al. 1995)
and WABS model (Morrisson \& McCammon 1983). To fit the spectra, XSPEC version 12.6.0 (Arnaud 1996) is used, and the model is
\begin{equation}
Model_1=Wabs(n_H)\times Mekal(T,z,A,norm).
\end{equation}
We fixed the redshift $z$ to 0.2523 and $n_H$
to the Galactic absorption 2.24$\times$10$^{20}$ cm$^{-2}$ (Dickey \& Lockman
1990). The fitting results are listed in Table 1.

\begin{table*}
\caption{Best-fit free parameters of Abell 1835: temperature $T$, abundance $A$, and normalized
constant $norm$.
The errors represent a
confidence level of 90\%.}
\begin{center}
{\footnotesize
\begin{tabular}{c@{\hspace{0.4cm}}c@{\hspace{0.4cm}}c@{\hspace{0.4cm}}
c@{\hspace{0.4cm}}c@{\hspace{0.4cm}}c@{\hspace{0.4cm}}c@{\hspace{0.4cm}}c}
\hline
\hline
\\
Annulus ($'$)   & $Temperature$& $A$ &$norm$ &$L_x$ (0.5-7.0 keV)&$\chi^{2}_{red}/d.o.f.$\\
 $r_1-r_2$& keV& solar &$10^{-3}$cm$^{-5}$ &$10^{45}$ergs$^{-1}$& \\
\hline\\
$0.0-0.25$ & $5.42 ^{+0.11}_{-0.11}$ & $0.40^{+0.04}_{-0.04}$&$5.92^{+0.07}_{-0.07}$&$1.03$&$0.986/368$\\
\\
$0.25-0.75$ & $6.60 ^{+0.14}_{-0.14}$ & $0.39^{+0.03}_{-0.03}$&$10.70^{+0.09}_{-0.09}$&$1.96$&$1.125/323$\\
\\
$0.75-1.50$& $9.85 ^{+0.82}_{-0.62}$ & $0.21^{+0.09}_{-0.09}$&$4.14^{+0.08}_{-0.08}$&$0.79$&$1.015/286$\\
\\
$1.50-2.25$ & $10.46 ^{+1.51}_{-1.30}$ & $0.24^{+0.20}_{-0.20}$&$2.22^{+0.09}_{-0.08}$&$0.42$&$0.868/157$\\
\\
$2.25-3.33$ & $11.96 ^{+4.48}_{-2.46}$ & $0.45^{+0.51}_{-0.45}$&$1.37^{+0.12}_{-0.10}$&$0.28$&$0.610/113$\\
\\
$3.33-6.00$ & $6.39 ^{+1.36}_{-1.08}$ & $0.36^{+0.33}_{-0.30}$&$1.61^{+0.11}_{-0.11}$&$0.20$&$1.234/125$\\
\\
\hline
\end{tabular}
}
\end{center}
\label{tablefit}
\end{table*}

The abundance is higher in the cluster center, which is understandable because the excess metal in the cluster center is produced in the cD galaxy and ejected into
the ICM (Makishima et al. 2001, Xiang et al. 2009).
From the deprojected temperature profile (squares in Fig.1), we see that the temperature decreases towards the center, which may be ascribed to the
gas cooling. Here we find the temperature profile is fitted well with the formula
\begin{equation}
T(r)=ae^{-(r-b)^2/2c^2}.
\end{equation}
The best-fit parameters are $a$ = 11.29 keV, $b$ = 2.78 arcmin,
and $c$ = 2.14 arcmin.
The best-fit profile is shown as a solid line in Fig.1.% Our calculations of mass and electron density are based on this profile.

To establish a comparison, we also obtained the temperatures from the projected spectra of {\it Chandra}, plotted as diamonds in Fig.1.
The average temperature excluding core ($r<0.75'$) is $\sim$9.6 keV,
which does not differ much from that of Markevitch (2002).  After deprojection, the temperatures do not differ much from the projected temperatures.
But they are still much higher than our new  result of {\it XMM-Newton}, $\sim$7.33 keV, which is consistent with that of Majerowicz et al. (2002) and Jia et al. (2004).
Therefore, even when the same method of data analysis is used, the temperature differences derived from {\it Chandra} and {\it XMM-Newton} data still exist.

\begin{figure}[ht]
\centerline{\psfig{file=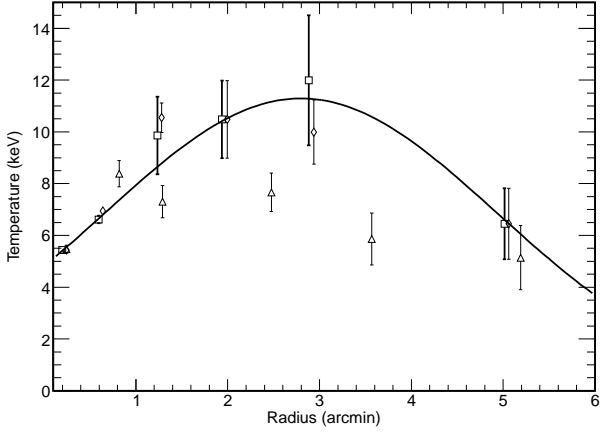,width=9cm}}
\caption{Temperature profiles of Abell 1835 with a confidence level of 90\%. Squares: deprojected temperatures of {\it Chandra}; diamonds: projected temperatures of {\it Chandra}; % from the single temperature
 triangles: deprojected temperatures of {\it XMM-Newton}.
The solid line is the best-fit profile of {\it Chandra} deprojected temperatures.}
\end{figure}

%%%%%%%%%%%%%%%%%%%%%%%%%%%%%%%%%%%%%%%%%%%%%%%%%%%%%%%%%%%%%%%%%%%%%%%%%%%%%%%
\section{Mass analysis}
In our mass model, we measure the spatially resolved radial temperature distribution from the deprojected spectra.
A double-$\beta$ model is adopted to fit the ICM density distribution.
\subsection{Electron density}
For calculating the electron density profile, we divided the cluster into 19 annular regions.
We calculated the deprojected photon counts in each shell. By using the deprojected abundance and the deprojected temperature
profile, we could estimate the normalized constant of each region, $norm$. Then we derive the deprojected
electron density $n_e$ of each region from Eq(3), shown as squares in Fig.2,

\begin{equation}
norm = \frac{10^{-14}}{4\pi [D_A(1+z)]^2}\int n_e n_H dV,
\end{equation}

where $D_A$ is the angular size distance to the source (cm) and $n_e$, $n_H$ (cm$^{-3}$) are the
electron and hydrogen densities, respectively.
We fitted the electron density profile with the double-$\beta$ model (Chen et al. 2003)
\begin{equation}
n_e(r)=n_{01} {\left[1+{\left(\frac{r}{r_{c1}}\right)}^2\right]}^{-{\frac{3}{2}}\beta_1}+n_{02}
{\left[1+{\left(\frac{r}{r_{c2}}\right)}^2\right]}^{-{\frac{3}{2}}\beta_2}.
\end{equation}
The best-fit parameters of {\it Chandra} are
$n_{01}=0.041\pm0.002 \textrm{ cm}^{-3}, $ $\beta_1=0.597\pm0.006,$ $ r_{c1}=0.325\pm0.013 \textrm{ arcmin},$
$n_{02}=0.114\pm0.003 \textrm{ cm}^{-3},$ $ \beta_2=12.81\pm3.080,$ $ r_{c2}=0.730\pm0.094 \textrm{ arcmin},$
$\chi^2_{red}=23.48,$ $d.o.f.=13.$\\

The best-fit profiles of $n_e$ from {\it Chandra} and {\it XMM-Newton} data are depicted in Fig.2,
which shows that the electron density derived from  {\it Chandra} is higher than that of  {\it XMM-Newton}.

\begin{figure}[ht]
\centerline{\psfig{file=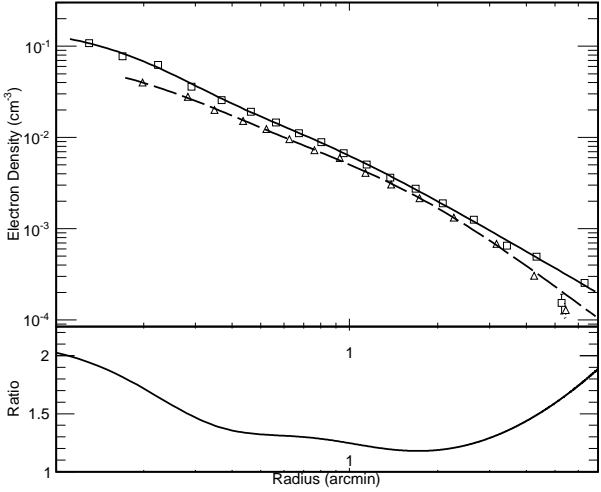,width=9cm}}
\caption{Top panel: electron density profiles of Abell 1835 from {\it Chandra} (squares) and {\it XMM-Newton} (triangles). The error bars are at a 68\%
confidence level. The solid line is the best-fit profile of {\it Chandra} results with a
double-$\beta$ model, and the dashed line for {\it XMM-Newton} results.
Bottom panel: the ratio between the electron density profiles of {\it  Chandra} and {\it XMM-Newton}.}
\end{figure}

\subsection{Mass calculation}
With the assumption of spherical symmetry and hydrostatic equilibrium, the total mass of cluster within
radius $r$ can be determined when the radial profiles of the gas density and temperature
are known. We calculated the gravitational mass of Abell 1835 with the
hydrostatic equation (Fabricant et al. 1980)
\begin{equation}
M_{tot}(<r)=-\frac{k_B T r^2}{G\mu
m_p}[\frac{d(\ln{n_e})}{dr}+\frac{d(\ln{T})}{dr}],
\end{equation}
where $k_B$ is the Boltzmann constant, $G$ is the gravitational constant, $\mu$
is the mean molecular weight of the gas in units of $m_p$, and $m_p$ is the proton mass.
For a fully ionized gas with a standard cosmic abundance,
a suitable value is $\mu$ = 0.60.

Using the best-fit profiles of the electron density $n_e(r)$ and the
deprojected temperature $T(r)$, we could obtain the total mass profile, as shown in
Fig.3. The total masses within the radius of $6'$ are $M_{tot-Chandra}=1.36\pm0.42\times10^{15}$ M$_{\odot}$ and $M_{tot-XMM}=0.84\pm0.10\times10^{15}$ M$_{\odot}$.
This shows that the {\it Chandra} result is much higher than the {\it XMM-Newton} result.
At the same time, the virial mass of {\it Chandra} is $M_{200}=2.0^{+0.4}_{-0.5}\times 10^{15}$ M$_{\odot}$ with $r_{200}=1.99$ Mpc, which is consistent with the result of Schmidt et al. (2001) when using the same cosmology model.

\begin{figure}[ht]
\centerline{\psfig{file=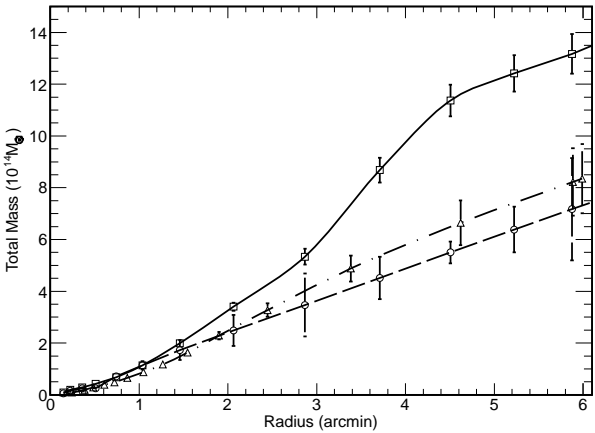,width=9cm}}
\caption{Total mass profiles of Abell 1835 from {\it XMM-Newton} (triangles), {\it Chandra} (squares)
and a method in subsection 5.2 (circles) fitting {\it Chanda} spectra with {\it XMM-Newton} temperatures.}
\end{figure}
\subsection{Mass within the optical lensing arc}
Allen et al. (1996) announced that there  is a lensing arc inside the cluster center.
 Richard et al. (2010) accurately recalculated the mass inside the radius of 250 kpc, $M_{SL}=2.83\pm0.41\times10^{14}$ M$_{\odot}$, based on seven multiply imaged systems under $\Lambda$CDM cosmology.
We estimated the X-ray projected mass inside the radius $M_{arc-Chandra}=1.85\pm0.20\times10^{14}$ M$_{\odot}$, $M_{arc-XMM}=1.40\pm0.15\times10^{14}$ M$_{\odot}$. The results, shown in Fig. 4,
are lower than the strong lensing mass derived by Richard et al. (2010).
The difference may be due to the assumption of spherical symmetry and hydrostatic equilibrium in the X-ray mass calculation (Gavazzi 2005).

The mass analysis of strong lensing cluster may be affected by orientation biases (Meneghetti et al . 2010b, 2011; Hennawi et al. 2007; Oguri \& Blandford 2009; Zitrin et al. 2011).
For Abell 1835, Corless et al. (2009) showed that the assumption of spherical symmetry is less problematic and that the orientation bias is weaker for this halo because of its smaller Einstein radius.
Thus, the mass estimates of Abell 1835 mentioned above are reasonable.
The radio plasma may provide some additional pressure to support the X-ray gas, so the assumption of hydrostatic equilibrium may be inaccurate.
\begin{figure}[ht]
\centerline{\psfig{file=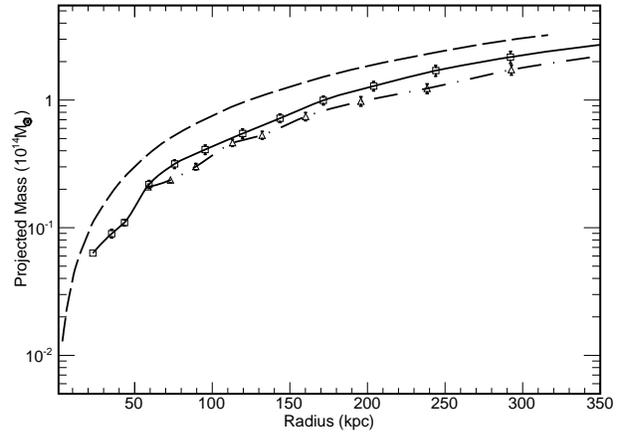,width=9cm}}
\caption{Projected mass profiles within 250 kpc of Abell 1835 from {\it Chandra} (squares) and {\it XMM-Newton} (triangles). The dashed line presents
the NFW profile of strong lensing mass of Richard et al. (2010).}
\end{figure}
\subsection{Gas mass fraction}
In galaxy clusters, gas , which has a temperature of a few keV, is an important component.
From the electron density, we can calculate gas mass and then the gas mass fraction, defined
as $f_{gas}(r)=M_{gas}(r)/M_{total}(r)$.

\begin{figure}[ht]
\centerline{\psfig{file=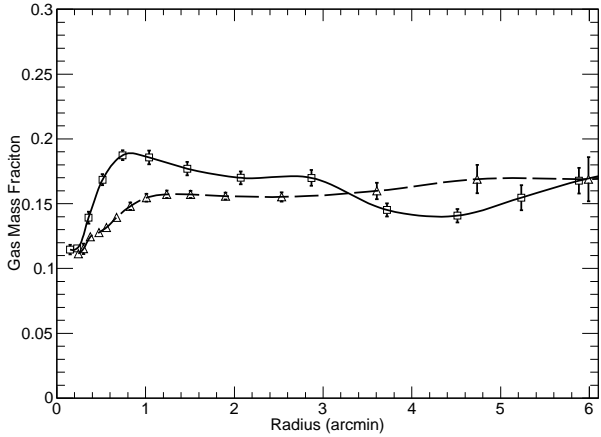,width=9cm}}
\caption{Gas mass fraction profiles of Abell 1835 from {\it Chandra} (squares) and {\it XMM-Newton} (triangles).}
\end{figure}
 Fig.5 shows the gas mass fraction of Abell 1835,
which drops clearly in the region $r<0.6'$, while keeping approximately constant in the outer region.
This indicates that the dark matter is more condensed than the gas in the central region of the cluster.

%%%%%%%%%%%%%%%%%%%%%%%%%%%%%%%%%%%%%%%%%%%%%%%%%%%%%%%%%%%%%%%%%%%%%%%%%%%%%%%

\section{Discussion}
Although we used the same method to analyse the data of Abell 1835 observed by {\it XMM-Newton} and {\it Chandra}, the temperature discrepancy still exists.
The average temperatures of Abell 1835 (excluding cool core) derived by {\it Chandra} and {\it XMM-Newton} are 9.6 keV and 7.33 keV, respectively.
in the following, we discuss the possible reasons for this.
\subsection{PSF scatter}
The effect of PSF of {\it pn/XMM-Newton} and {\it MOS/XMM-Newton} is larger than that of {\it ACIS/Chandra}.
PSF may scatter a significant fraction of the adjacent regions' emission into
the regions that we analyze. 

Applying a combined direct demodulation and deprojection technique to subtract the effects of PSF, Wang et al. (2010) found that
the central electron density increases by 30\%, while the temperature profile is similar to the result using only a deprojection
method.
They also estimated that the effects of PSF were not important for the temperature profile.
 Nevalainen et al. (2010) pointed out that PSF scattering may cause differences in temperature measurements with different instruments of
a maximum of 2\%, which is negligible compared to the statistical uncertainties with the minimum width of rings set to 1.5$'$.
Therefore, PSF is not a primary reason for the temperature discrepancy.

\subsection{Discrepancy between the temperatures}

Since the effects of PSF and deprojection technology on temperature are not the primary reasons for the temperature discrepancy,
the difference of the calibration of {\it XMM-Newton} and {\it Chandra} may lead to the temperature discrepancy.
Nevalainen et al. (2010) found that the difference of the calibration of these two instruments is about 10-15\% on average and the maximum difference is $\sim25\%$. %  The ratio of the average temperature of A1835 is in this region.
Therefore, it may be the main factor of the temperature discrepancy of Abell 1835.

In order to contrast the temperatures of Abell 1835 determined by {\it Chandra} and {\it XMM-Newton}, we tried to fit {\it Chandra} and {\it XMM-Newton} spectra with temperatures fixed to the other one's temperatures.
  Because the temperatures of  {\it XMM-Newton} and {\it Chandra} are significantly different in annulus $0.75'-3.33'$,
  we combined and rebinned the three spectra of $0.75'-3.33'$. The fitting results are listed in Table 2.
It seems that {\it Chandra} spectra can be fitted well with {\it XMM-Newton} temperature, while {\it XMM-Newton} spectra can not fit well with {\it Chandra} temperature.
 As a result, we think that {\it XMM-Newton} has a stronger restriction on temperature.

 \begin{table*}
\caption{Best-fit parameters of $0.75'-3.33'$: $temperature$, $abundance$, and normalized
constant $norm$.
The errors represent a confidence level of 90\%.}
\begin{center}
{\footnotesize
\begin{tabular}{l@{\hspace{0.5cm}}c@{\hspace{0.5cm}}c@{\hspace{0.5cm}}
c@{\hspace{0.5cm}}c@{\hspace{0.5cm}}c@{\hspace{0.5cm}}c@{\hspace{0.5cm}}c}
\hline
\hline
\\
Parameter&{\it Chandra} &  {\it Chandra}&  {\it XMM-Newton}  & {\it XMM-Newton}
\\
&  & ($T=T_{\textrm {\tiny XMM}}$)&  &($T=T_{\textrm {\tiny Chandra}}$)
\\
\hline
\\
$Temperature$ (keV)  & $10.44^{+0.72}_{-0.72}$ & $7.73$ (fix)& $7.73^{+0.43}_{-0.34}$  &10.44 (fix)
\\
\\
$Abundance$ (solar)  & $0.27^{+0.09}_{-0.09}$&0.27 (fix)   &$ 0.31^{+0.09}_{-0.09}$&0.31 (fix)
\\
\\
 $norm $ ($10^{-3}$cm$^{-5}$)& $4.17^{+0.07}_{-0.07}$&  $4.16^{+0.08}_{-0.08}$& $4.34^{+0.06}_{-0.06}$  & $4.34^{+0.06}_{-0.06}$
 \\
\\
$\chi^{2}_{\mathrm{red}}$/$d.o.f.$ & 0.86/191 &1.24/193 &0.98/165  &1.51/167
\\
\\
$Probability$ &$0.91$  & 0.013&0.56 & $2.24\times10^{-5}$
\\
\\
\hline
\end{tabular}
}
\end{center}
\label{tablefit}
\end{table*}

Furthermore, we calculated the total mass with {\it XMM-Newton} temperatures and {\it Chandra} electron density based on {\it XMM-Newton} temperatures (hereafter $M_{Chandra (T=T_{XMM})}$),
which is $0.74\pm0.05 \times 10^{15}$M$_{\odot}$ within the radius of $6.0'$ (circles in Fig.3).
It is similar to the total mass derived from {\it XMM-Newton} data.

To compare with the weak lensing mass, we also calculated the projected masses at $r_{2500}$, $r_{1000}$ and $r_{500}$.
The results are listed in Table 3,
which shows that the weak lensing mass is lower than $M_{Chandra}$, and a little higher than $M_{XMM}$ and $M_{Chandra (T=T_{XMM})}$ at $r_{500}$.

Because of the non-hydrostatic state and non-equilibrium processes in clusters, the X-ray masses may be underestimated (e.g., Rasia et al. 2004; Piffaretti et al. 2004; Nagai et al. 2007).
In Table 4, we list the comparison of the X-ray mass and the weak lensing mass from both the observational  and simulated analyses.
It shows that the X-ray mass is always consistent with, or a little smaller than, the weak lensing mass (Zhang et al. 2008; Mahdavi et al. 2008; Zhang et al. 2010; Meneghetti et al. 2010; Rasia et al. 2012).
Consequently, our $M_{XMM}$ and $M_{Chandra (T=T_{XMM})}$ are more suitable.

With {\it XMM-Newton} temperatures, we can get more reliable masses from {\it XMM-Newton} and {\it   Chandra} data. Because {\it XMM-Newton} has a stronger restriction on temperature,
the temperatures derived from {\it XMM-Newton} data may be more reliable.

\begin{table*}
\caption{Projected masses of Abell 1835. The weak lensing masses are reported in Zhang et al. (2010). $r_{2500}$, $r_{1000}$ and $r_{500}$ are determined from the weak lensing analysis.}
\begin{center}
{\footnotesize
\begin{tabular}{c@{\hspace{0.5cm}}c@{\hspace{0.5cm}}c@{\hspace{0.5cm}}
c@{\hspace{0.5cm}}c@{\hspace{0.5cm}}c@{\hspace{0.5cm}}c@{\hspace{0.5cm}}c}
\hline
\hline
\\

$R$&$M_{{\it Chandra}}$&$M_{{\it XMM-Newton}}$&$M_{{\it Chandra}(T=T_{\textrm{\tiny XMM}})}$&$M_{wl}$
\\
&$10^{14}$ M$_{\odot}$&$10^{14}$ M$_{\odot}$&$10^{14}$ M$_{\odot}$&$10^{14}$ M$_{\odot}$
\\
\hline
\\
$r_{2500}$&$4.42\pm0.52$&$3.15\pm0.62$&$3.28\pm0.40$&$2.88\pm0.58$
\\
\\
$r_{1000}$&$7.11\pm1.86$&$5.80\pm1.42$&$5.28\pm1.39$&$6.15\pm0.95$
\\
\\
$r_{500}$&$13.80\pm3.91$&$8.99\pm3.60$&$8.19\pm3.21$&$9.65\pm1.70$
 \\
\\
\hline
\end{tabular}
}
\end{center}
\label{tablefit}
\end{table*}

\begin{table*}
\caption{Ratio between X-ray and weak lensing masses.}
\centering
\begin{tabular}{clccc}
\hline
\hline
&Sample&&$M_X/M_{WL}$\\
 && $R_{2500}$ & $R_{1000}$& $R_{500}$\\
\hline
\\
Observation&Zhang et al. 2010 (unrelaxed)            &   1.00 $\pm$ 0.07  & 0.97  $\pm$ 0.05  & 0.99  $\pm$ 0.07 \\
\\
&Zhang et al. 2010 (relaxed)       &   1.04 $\pm$ 0.08  & 0.96 $\pm$ 0.05   & 0.91 $\pm$ 0.06 \\
\\
&Mahdavi et al. 2008                            & 1.03 $\pm$ 0.07 & 0.90 $\pm$ 0.09 & 0.78$\pm$ 0.09 \\

\\
\hline
\\
Simulation&Rasia et al. 2012                                        &  0.83 $\pm$ 0.02 & 0.80 $\pm$ 0.02 & 0.75 $\pm $0.02 \\
\\
&Meneghetti et al. 2010                                    &  0.90 $\pm $0.04 & 0.86 $\pm$ 0.02 &  0.88$ \pm $0.02 \\
\\
\hline
\end{tabular}
\label{tab:compa}
\end{table*}
%%%%%%%%%%%%%%%%%%%%%%%%%%%%%%%%%%%%%%%%%%%%%%%%%%%%%%%%%%%%%%%%%%%%%%%%%%%%%%%
\section{Conclusion}
We have presented a detailed analysis of about 117 ksec of {\it Chandra}
observations on the galaxy cluster Abell 1835. Through the deprojected spectra
analysis, we derived the deprojected temperatures, which do not differ much from the temperatures without deprojection and
are still much higher than those of {\it XMM-Newton}.

The total mass within the radius of $6'$, $M_{tot-Chandra}$ = 1.36$\pm$0.42$\times$10$^{15}$ M$_{\odot}$, is much higher than the total mass of
$M_{tot-XMM}$ = 0.84$\pm$0.10$\times$10$^{15}$ M$_{\odot}$ derived from {\it XMM-Newton} data.
 The difference of total mass is due to the discrepancy of the temperatures derived from {\it Chandra} and {\it XMM-Newton}.
We also calculated the projected mass within the optical lensing arc, which is lower than the strong lensing mass derived by Richard et al. (2010).

After deprojection and PSF correction (Wang et al. 2010), the temperature and total mass of Abell 1835 resulting from {\it Chandra} are still different from those of  {\it XMM-Newton}.
These differences may result from the calibration of the two instruments.
We find that {\it XMM-Newton} has a stronger restriction on temperature.
And with {\it XMM-Newton} temperatures, the projected mass from {\it Chandra} data is lower than the weak lensing mass and consistent with other observational analyses.
For these reasons, the temperatures obtained from {\it XMM-Newton} may be more reliable.
%%%%%%%%%%%%%%%%%%%%%%%%%%%%%%%%%%%%%%%%%%%%%%%%%%%%%%%%%%%%%%%%%%%%%%%%%%%%%%%
\begin{acknowledgements}
We would like to thank the referee for insightful and helpful comments, which
improved the paper significantly.
This research was supported by Project 11003018, sponsored by National Nature Science Foundation of China.
\end{acknowledgements}

%%%%%%%%%%%%%%%%%%%%%%%%%%%%%%%%%%%%%%%%%%%%%%%%%%%%%%%%%%%%%%%%%%%%%%%%%%%%%%%


\begin{thebibliography}{}


\bibitem{} Andersson K.E., \& Madejski G.M., 2004, ApJ, 607, 190
\bibitem{} Allen S.W., Fabian A.C., Edge A.C., et al., 1996, MNRAS, 283, 263
\bibitem{} Allen S.W., 2000, MNRAS, 315, 269
\bibitem{} Bartelmann M., 1995, A\&A, 299, 11
\bibitem{} B\"{o}hringer H., Matsushita K., Churazov E., et al., 2002, A\&A, 382, 804
\bibitem{} Chen Y., Ikebe Y., \& B\"{o}hringer H., 2003, A\&A, 407, 41
\bibitem{} Clowe D., \& Schneider P., 2002, A\&A, 395, 385
\bibitem{} Corless V.L., King L.J., \& Clowe D., 2009, MNRAS, 393, 1235
\bibitem{} Fabian A.C., 1988, Sci, 242, 1586
\bibitem{} Fabricant D., Lecar M., \& Gorenstein P., 1980, ApJ, 241, 552
\bibitem{} Gavazzi R., 2010, A\&A, 443, 793
\bibitem{} Hennawi J. F., Dalal N., Bode P., \& Ostriker, J. P. 2007, ApJ, 654, 714
\bibitem{} Jia S.M., Chen Y., Lu F.J., et al., 2004, A\&A, 423, 65
\bibitem{} Jia S.M., Chen Y., \& Chen L., 2006, ChJAA, 6, 181
\bibitem{} Kaastra J.S., 1992, An X-ray Spectral Code for Optically Thin Plasma
(Internal Sron-Leiden Report, updated version 2.0 )
\bibitem{} Lemze D., Barkana R., Broadhurst T.J., et al., 2008, MNRAS, 386, 1092
\bibitem{} Liedahl D.A., Osterheld A.L., \& Goldstein W.H., 1995, ApJL, 438, 115
\bibitem{} Loeb Abraham, \& Mao Shude, 1994, ApJ, 435, 109
\bibitem{} Majerowicz S., Neumann D.M., \& Reiprich T.H., 2002, A\&A, 394, 77
\bibitem{} Makishima K., Ezawa H., Fukuzawa Y., et al., 2001, PASJ, 53, 401
\bibitem{} Markevitch M., 2002, preprint astro-ph/0205333
\bibitem{} Markevitch M., Bautz M.W., Biller B., et al., 2003, ApJ, 583, 70
\bibitem{} Mewe R., Gronenschild E.H.B.M., \& van den Oord G.H.J., 1985, A\&AS, 62, 197
\bibitem{} Mewe R., Lemen J.R., \& van den Oord G.H.J. 1986, A\&AS, 65, 511
\bibitem{} Mahdavi A., Hoekstra H., Babul A., et al., 2008, MNRAS, 384, 1567
\bibitem{} Meneghetti M., Rasia E., Merten J., et al., 2010, A\&A, 514, 93
\bibitem{} Meneghetti M., Fedeli C., Pace F., et al., 2010, A\&A, 519, 90
\bibitem{} Meneghetti M., Fedeli C., Zitrin A., et al., 2011, A\&A, 530, 17
\bibitem{} Morandi A., Marceau L., Jack S., et al., 2011, [arXiv:1111.6189]
\bibitem{} Morrison R., \& McCammon D., 1983, ApJ, 270, 119
\bibitem{} Nagai D., Vikhlinin A., \& Kravtsov A. V., 2007, ApJ, 655, 98
\bibitem{} Nevalainen J., David L., \& Guainazzi M., 2010, A\&A, 523, 22
\bibitem{} Nulsen P. E. J., \& B\"{o}hringer H., 1995, MNRAS, 274, 1093
\bibitem{}	Oguri M., \& Blandford R. D., 2009, MNRAS, 392, 930
\bibitem{} Piffaretti R., Kaastra J., Tamura T., et al., 2004, ogci.conf, 131
\bibitem{} Rasia E., Tormen G., \& Moscardini L., 2004, MNRAS, 351, 237
\bibitem{} Rasia E., Meneghetti M., Martino R., et al., 2012, [arXiv:1201.1569]
\bibitem{} Richard J., Smith G.P., Kneib J.P., et al., 2010, MNRAS, 404, 325
\bibitem{} Schmidt R.W., Allen S.W., \& Fabian A.C., 2001, MNRAS, 327, 1057
\bibitem{} Smith G. P., Kneib J.-P., Smail I., et al., 2005, MNRAS, 359, 417
\bibitem{} Wang Y.S., Jia S.M., \& Chen Y. 2010, SCIENCE CHINA Physics, Mechanics \& Astronomy, 53, 183
\bibitem{} Xiang F., Rudometkin E., Churazov E., et al., 2009, MNRAS, 398, 575
\bibitem{} Zhang Y.Y., Finoguenov A., B\"{o}hringer H., et al., 2008, A\&A, 482, 451
\bibitem{} Zhang Y.Y., Okabe N., Finoguenov A., et al., 2010, ApJ, 711, 1033
\bibitem{} Zitrin A., Broadhurst T., Coe D., et al., 2011, ApJ, 742, 117
\end{thebibliography}
\end{document}